\documentclass{article}

\usepackage{graphicx}
\usepackage{amssymb}
\usepackage{epstopdf}
\usepackage{amsmath}
\usepackage{natbib}
\usepackage{subfigure}

\def\pd#1#2{\frac{\partial #1}{\partial #2}}

\def\\tg{\mathop{\rm tg \nolimits}}

\begin{document}         

{\bf Two Kuril tsunamis and analytical long wave theory}  
\bigskip

I. M. Mindlin 

State Technical University of Nizhny Novgorod, Russia 

E-mail address: ilia.mindlin@gmail.com
\bigskip

This paper addresses long  waves on the water surface.
It is assumed that initially the water surface has not yet been displaced 
from its mean level, but the velocity field has already become different from zero. This means that the motion of a body of water is triggered by a sudden change in the velocity field. 
 
The long wave is modelled mathematically as a specific wave packet. The model is used to estimate  duration of the wave origin formation, size of the origin,  water elevation in the origin, energy supplied to the water by the quake, distribution of wave heights in the wave 
packet. Relationship between group velocity of the packet and phase velocity 
of the packet's wave of maximum height is revealed. 
 These results are applied to  Kuril tsunamis of November 2006 and 
January 2007.
\bigskip

{\bf Motivation of the study}
\bigskip
 
 Numerical study of two tsunamis triggered by the earthquakes of the 2006 and 2007 near the Kuril Islands were prsented in [1]. As to comparison of the tsunamis, one can read in [1]: 

1. "The dimensions of the 2007 tsunami source were smaller than those of the 2006 event, ...the 2007 tsunami had higher dominant frequency [1, p.115].

2. " The wave energy of the 2007 Kuril tsunami is reduced compared to that of 2006 Kuril tsunami�" [1, p.115].

3. "at remout sites ...the ratio of 2006/2007 far-field wave heights is typically around 3:1" [1, p.115].

These three assertions, especially with the ratio 3:1, stimulate the following 
analytical study of gravitational water waves.
\bigskip

{\bf 1. Problem outline. }
\bigskip                      

In the present study the deep-water waves are considered
which start to propagate away from an initially disturbed body of water. 
Then the water is acted on by no external force other than gravity. 
It is assumed that the free surface of the water is infinite in extent and 
the pressure along the surface is constant.

Referring to figure 1 and  assuming that the motion is two-dimensional, 
in the $(x,y)$-plane,
consider flow of an ideal heavy uniform liquid of density 
$\gamma$; the $x$-axis is oriented upward and the $y$-axis in horizontal
direction. 

Let the curve $\,\,\Gamma\,\,$ (in figure 1) be the trace of the free
surface $\,\,S\,\,$ in the $\,\,(x,y)\,\,$ plane,
$\,x=f<0,\,$ $y=0$ be the coordinates of the pole $\,O_1\,$ of the polar
coordinate system in the $\,(x,y)\,$ plane,
$\,\theta\,$ be the polar angle measured from the positive $x$-axis
in the counterclockwise direction, $\,t\,$ be the time. 
The external pressure, $\,P_*\,$, on the free surface is constant.
The liquid fills the space below the free surface.
The equilibrium position of the free surface is horizontal plane $x=0$.
\begin{figure}
	\resizebox{\textwidth}{!}
		{\includegraphics{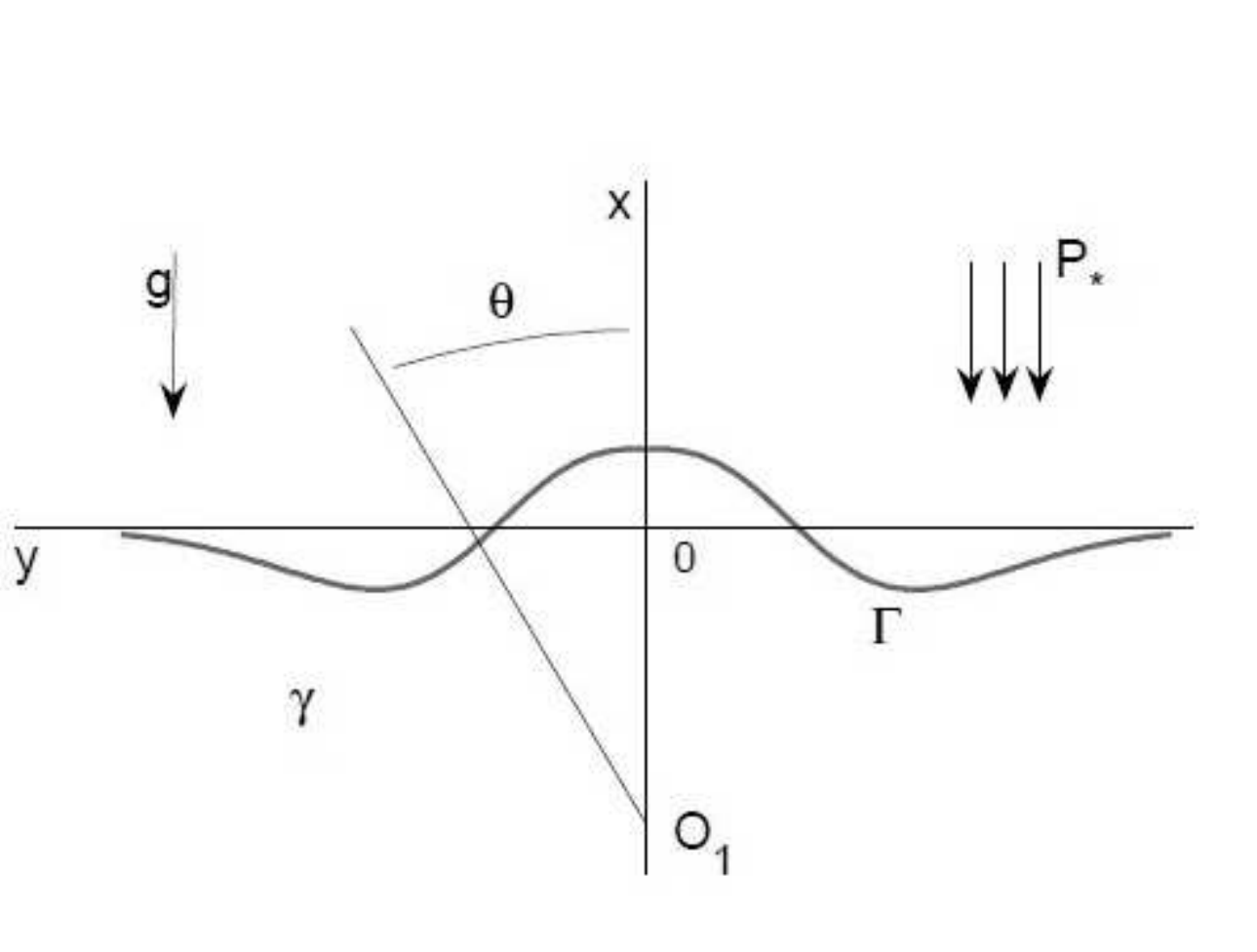}} 
	\caption{
	Coordinate systems and sketch of the free surface of a liquid. 
	}
	\label{qu-.1-.10}
\end{figure} 
    Assuming that the waves are generated by 
an initial disturbance to the water and the horizontal dimensions 
of the initially disturbed body of the water are much larger than 
the magnitude of the water surface displacement in the wave origin, 
equations for the water surface displacement has been obtained 
in parametric form [2]
$$
x=cW(\theta, t),\,\,\,\,\,\,\,y=(x-f)\tan\theta,     \eqno(1.1)
	$$
$$
W=\sum\limits_{n=1}^{+\infty}a_nT_n(\theta,\tau),\,\,\,\,\,\,\, 
  -\frac{\pi}{2}<\theta<\frac{\pi}{2},
	$$
Formulas for $T_n(\theta,\tau)$ are given  in [2], 
the constants $a_n$ determine initial displacement to the free surface 
and initial velocity field.  

Formally, equations (1.1) for each specified function
$\,\,W\,\,$ describe a family of curves depending on $\,f,\,$ $\,t\,$
 being considered as constant. The value  of $\,\,f\,\,$ 
determines the horizontal scale of the problem. 
Though the function $W(\theta,t)$ is a linear combination of
the functions $T_n(\theta,\tau)$,  the waves (1.1) do not obey the 
principle of linear superposition: the implicit form of the waves 
is \linebreak $x=cW_0(\hbox{arc tangent}(y/(x-f)),t)$. 

This means that the free surface waves  produced by 
the initial disturbance to the water is a nonlinear mixture 
 of finite or infinite (it depends on initial conditions) set of the specific wave packets 
$$
x=cT_n(\theta, t),\,\,\,\,\,\,\,y=(x-f)\tan\theta.
	$$
The  wave packets of different numbers travel at different speed,  and evolution of each packet in the mixture is not influenced by evolution of the others. 

\begin{figure}
	\resizebox{\textwidth}{!}
		{\includegraphics{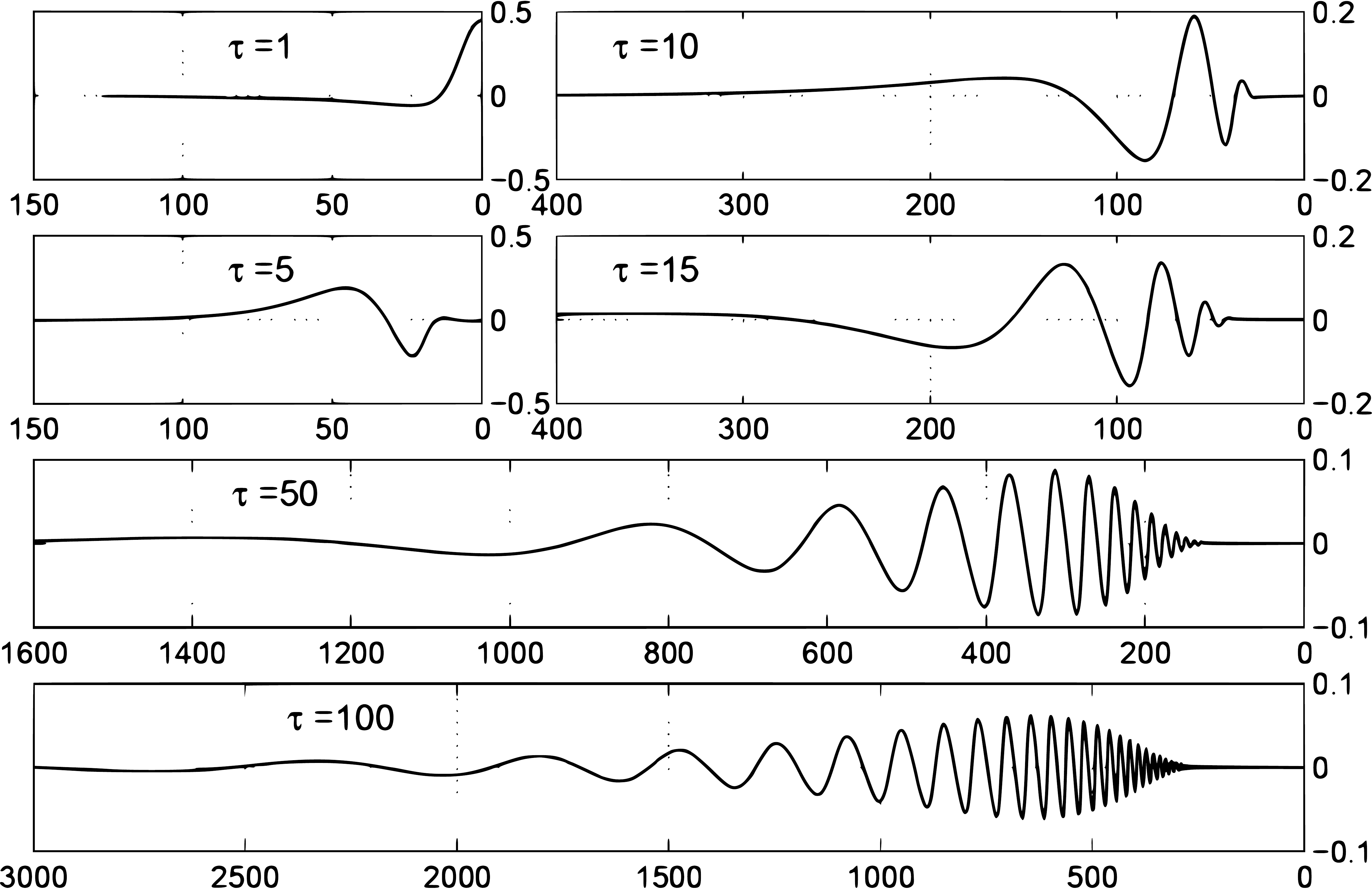}} 
	\caption{
Profiles of the packet (2.1) at different values of $\tau$ shown in the figure near the corresponding curves, $c=0.5,\,\,\,f=-10$.  	
	}
	\label{qu-.1-.10}
\end{figure} 
\bigskip

{\bf 2. Theoretical model for long water  waves}
\bigskip

The surface waves are modelled by a specific wave packet 
$$
x=c\,\frac{1}{\sqrt{2|f|}}T_1(\theta,\tau),\,\,\,\, y=(x-f)\tan\theta, \,\,\,\,\,\,
-\pi/2< \theta < \pi/2             \eqno(2.1)
	$$
$$
T_1(\theta,\tau)=\int\limits_0^{+\infty}
x^2e^{-x^2/2}\cos\left(\frac{1}{2}x^2\tan\theta\right)\sin(\tau x)\,dx,
\, \,\,\,\,\,t=\tau\,{\sqrt{2|f|}}  .   
	$$
By the moment $t=0$ the free surface has not yet been displaced 
from its mean level (the horizontal plane $x=0$), but the velocity field has 
already become different from zero [2]: 
$$
 T_1(\theta, 0)=0,\,\,\,\,\,\,\,\,
\left.\pd{T_1}{\tau}\right|_{\tau=0}=
2\cos^2\theta\cdot\cos(2\theta)
	$$
Figure 2 displays profiles of the packet (2.1) at $\tau=1,\,5,\,10,\,15,\,50,\,100.$

The wave packet (2.1) travels faster then any other specific packet  
and with time leaves behind the other packets. 

The model is adopted as a rough mathematical model for 
propagation of long waves through an open sea, and is not 
intended for detailed quantitative description of the tsunamis referred to below.

All equations are written in non-dimensional variables.
Since the problem has no characteristic linear size, 
the dimensional unit of length, $\,\,L_*,\,\,$ is a free parameter.
But for applications in the section 6, the value of $L_*$,    
as well as the value of $|f|L_*$, will be obtained  from instrumental data.
The dimensional unit of time, $\,\,T_*,\,\,$ is defined
by the relation $\,\,T_*^2g=L_*\,$, where $\,\,g\,\,$ is the acceleration
of free fall. The non-dimensional acceleration of free fall is equal to unity.
All parameters, variables and equations are made non-dimensional by 
the quantities $\,L_*,\,T_*,\,P_* $ and 
the density of water $\gamma_*=1000\,$ $\hbox{kg/m}^3$
\bigskip

{\bf 3. Properties of the specific wave packet }
\bigskip

Zeros of the wave packet (2.1)  are defined by the equation $T_1(\theta,\tau)=0$ and 
(at fixed value of $\tau$) are situated in the numbered rays 
$\theta=\theta_k(\tau)$ ($k$ is the number of a zero). 
Zeros $\theta_k(\tau)$ of the wave packet are independent of $f$. 

The term 'wave'\, means a
 section of the packet which is singled out by three consecutive zeros and consists of a crest and the trough following or preceding the crest.
The level difference between the crest and the trough  
is referred to as the wave height (or height of the wave), 
and the distances between two successive zeros is "half-wave-length". 

At any particular moment of time, each specific wave packet contains only one 
wave of maximum height (WMH) on semiaxis $y>0$  
(the situation with two waves of equal maximum height can be ignored), so
the zeros of WMH constitute a 'natural frame of reference' for other zeros:

Let $\theta_r(\tau)$ and $\theta_f(\tau)$  denote two of the three zeros 
of the WMH which correspond to minimum and maximum of the three 
$|\tan\theta(\tau)|$ respectively. By the zero $\theta_f(\tau)$ we define 
the front of WMH at the instant $\tau$, by $\theta_r(\tau)$ the rear 
of the wave is determined.  

Horizontal coordinate of the zero $\theta_k(\tau)$ is given 
by $y_k=-f\cdot \tan\theta_k(\tau)$ and the function 
$T_1(\theta,\tau)$ is 
independent of $f$. This leads to the following 

{\it Assertion:} At any given value of $\tau$

i) for any wave of the packet the quantity $\Delta(\tau)=h(\tau)\sqrt{2|f|}$
($h(\tau)$ is the height of the wave) is independent of $f$; 

ii) the ratio of the distances $y_{k+1}(\tau)-y_k(\tau)$ and
$y_k(\tau)-y_{k-1}(\tau)$ between any successive zeros of the wave packet  
(and, consequently, the ratio of the lengths of two successeive waves) is
independent of $f$.

Let $L_*\,\,$ be the dimensional unit of length (in metres), then $T_*=\sqrt{L_*/g}\,\,$   is the dimensional unit of time (in seconds).

At an instant $t_*$ the dimensional coordinate of any zero 
$\theta=\theta(\tau)$ is 
$$
y_*(t_*)=|f|L_*\cdot \tan\theta(\tau)\cdot 10^{-3}\,\,\hbox{km},\,\,\,\,\,\,
t_*=\tau\sqrt{2|f|}\cdot T_*/60\,\,\hbox{min},
	$$
and, consequently, the ratio 
$$
\lambda(\tau)=\frac{y_*(t_*)}{gt_*^2}=
\frac{\tan\theta(\tau)}{2\tau^2}               \eqno(3.1)
	$$
depends only on $\tau$.

When the distance is measured in kilometres and time in minutes, it is conveniet to rewrite formula (3.1) as
$$
\lambda_*(\tau)=\frac{y_*(t_*)}{t_*^2}\,\hbox{km/min$^2$}=
17.64\,\frac{\tan\theta(\tau)}{\tau^2}\,\hbox{km/min$^2$}.      \eqno(3.2)
	$$

 Given a fixed value of $\tau$, $\tan\theta(\tau)$  can be calculated from equations 
(2.1) of the wave packet, and corresponding value of $\lambda_*(\tau)$ can be obtained from (3.2). 
 
Table 1 shows computed characteristics of the WMH:
$\tan\theta_f,\,$ $\lambda_*(\tau),\,$ and 
$\Delta(\tau)=H(\tau)\sqrt{2|f|}\,\,$ ($\,\,cH(\tau)$ is the maximum wave height)
obtained from (2.1) and (3.2) at $c=1$ $f=-10$).

\begin{center}
TABLE 1. Computed characteristics of the wave of maximum height.
\end{center}
\begin{footnotesize}

\begin{center}
\begin{tabular}{|p{10mm}  |p{10mm}  |p{10mm}  |p{10mm}  |p{10mm}  | p{10mm} | p{10mm}| }
\hline $\tau$      & 90     & 100    & 110    & 120    & 130    & 140    \\
 $\tan\theta_f  $    & 58.000 & 65.695 & 70.734 & 78.421 & 83.470 & 88.559 \\                           
 $\lambda_*$         & 0.1263 & 0.1159 & 0.1031 & 0.0960 & 0.0871 & 0.0797 \\
 $\Delta(\tau)  $        & 0.3840 & 0.3661 & 0.3507 & 0.3371 & 0.3255 & 0.3152 \\

$\lambda_*\cdot\tau$ & 11.368 & 11.589 & 11.343 & 11.528 & 11.326 & 11.158 \\

 $u $              & 0.5026 & 0.7695 & 0.5039 & 0.7687 & 0.5049 & 0.7606  \\
\hline
\end{tabular}\\
\end{center}

\begin{center}
\begin{tabular}{|p{10mm}  |p{10mm}  |p{10mm}  |p{10mm}  |p{10mm}  | p{10mm} | p{10mm}| }
\hline $\tau$      & 150    & 160     & 170     & 180     & 200     & 220      \\
 $\tan\theta_f  $    & 96.203 & 101.294 & 108.935 & 116.607 & 126.767 & 142.072  \\                           
 $\lambda_*$         & 0.0754 & 0.0698  & 0.0665  & 0.0635  & 0.0559  & 0.0518   \\
 $\Delta(\tau) $         & 0.3050 & 0.2960  & 0.2884  & 0.2804  & 0.2676  & 0.2554   \\

$\lambda_*\cdot\tau$ & 11.313 & 11.168  & 11.304  & 11.427  & 11.181  & 11.391   \\

 $u $              & 0.7644 & 0.5091  & 0.7641  & 0.7672  & 0.5081  & 0.7652  \\
\hline
\end{tabular}\\
\end{center}

\begin{center}
\begin{tabular}{|p{10mm}  |p{10mm}  |p{10mm}  |p{10mm}  |p{10mm}  | p{10mm} | p{10mm}| }
\hline $\tau$      & 240     & 260     & 280     & 300     & 310     & 320     \\
 $\tan\theta_f  $    & 154.802 & 167.532 & 180.265 & 192.996 & 198.071 & 205.727 \\ 
 $\lambda_*$         & 0.0474  & 0.0437  & 0.0406  & 0.0378  & 0.0364  & 0.0354  \\
 $\Delta(\tau) $         & 0.2446  & 0.2367  & 0.2281  & 0.2204  & 0.2175  & 0.2141  \\

$\lambda_*\cdot\tau$ & 11.378  & 11.368  & 11.362  & 11.348  & 11.271  & 11.341  \\

 $u $              & 0.6365  & 0.6366  & 0.6366  & 0.6365  & 0.5075  & 0.7656  \\
\hline
\end{tabular}\\
\end{center}

\begin{center}
\begin{tabular}{|p{10mm}  |p{10mm}  |p{10mm}  |p{10mm}  |p{10mm}  | p{10mm} | p{10mm}| }
\hline $\tau$      & 330     & 340     & 360     & 370     & 380     & 390    \\
 $\tan\theta_f  $    & 210.802 & 218.458 & 231.189 & 236.267 & 243.920 & 249.002 \\ 
 $\lambda_*$         & 0.0341  & 0.0333  & 0.0315  & 0.0304  & 0.0298  & 0.0289  \\
 $\Delta(\tau) $         & 0.2108  & 0.2077  & 0.2019  & 0.1991  & 0.1965  & 0.1950  \\

$\lambda_*\cdot\tau$ & 11.268  & 11.334  & 11.328  & 11.264  & 11.323  & 11.262  \\

 $u $              & 0.5075  & 0.7656  & 0.6365  & 0.5078  & 0.7653  & 0.5082  \\
\hline
\end{tabular}\\
\end{center}

\begin{center}
\begin{tabular}{|p{10mm}  |p{10mm}  |p{10mm}  |p{10mm}  |p{10mm}  | p{10mm} | p{10mm}| }
\hline $\tau$      & 400     & 410     & 420     & 440     & 460     & 470      \\
 $\tan\theta_f  $    & 254.094 & 261.734 & 269.387 & 279.559 & 292.291 & 299.931  \\                           
 $\lambda_*$         & 0.0280  & 0.0275  & 0.0269  & 0.0255  & 0.0244  & 0.0239   \\
 $\Delta(\tau) $         & 0.1926  & 0.1902  & 0.1879  & 0.1836  & 0.1796  & 0.1776   \\
                                                                               
$\lambda_*\cdot\tau$ & 11.205  & 11.261  & 11.314  & 11.208  & 11.209  & 11.257   \\

 $u $              & 0.5092  & 0.7640  & 0.7653  & 0.5086  & 0.6366  & 0.7640  \\
\hline
\end{tabular}\\
\end{center}

\end{footnotesize}

In Table 1, values of the ratios $u=\Delta\tan\theta_f/\Delta\tau$ are given for each two neighbouring columns (for instance, for the columns $\tau=90$ and $\tau=100$ we find $u=(65.695-58.000)/(100-90)=0.7695$).

During a time interval $t$ min, the front of the WMH travels a distance 
$y(t)$ km at the average speed  
$$
V=\frac{y(t)}{t}=
\lambda_*(\tau)\cdot \tau \sqrt{2|f|}\cdot \frac{T_*}{60}\,\hbox{km/min}.        
                                              \eqno(3.3)  
	$$                                                            
The value of the average speed during time interval 
$\Delta \tau=\tau_{n+1}-\tau_n$ equals
$$
v=u\,\sqrt{|f|L_*g/2}\,\hbox{m/s},\,\,\,\,\,\,              
u=[\tan\theta_f(\tau_{n+1})-\tan\theta_f(\tau_n)]/\Delta\tau,    \eqno(3.4)
$$
where $\tau_n$ and $u$ are given in Table 1.
\begin{figure}
	\resizebox{\textwidth}{!}
		{\includegraphics{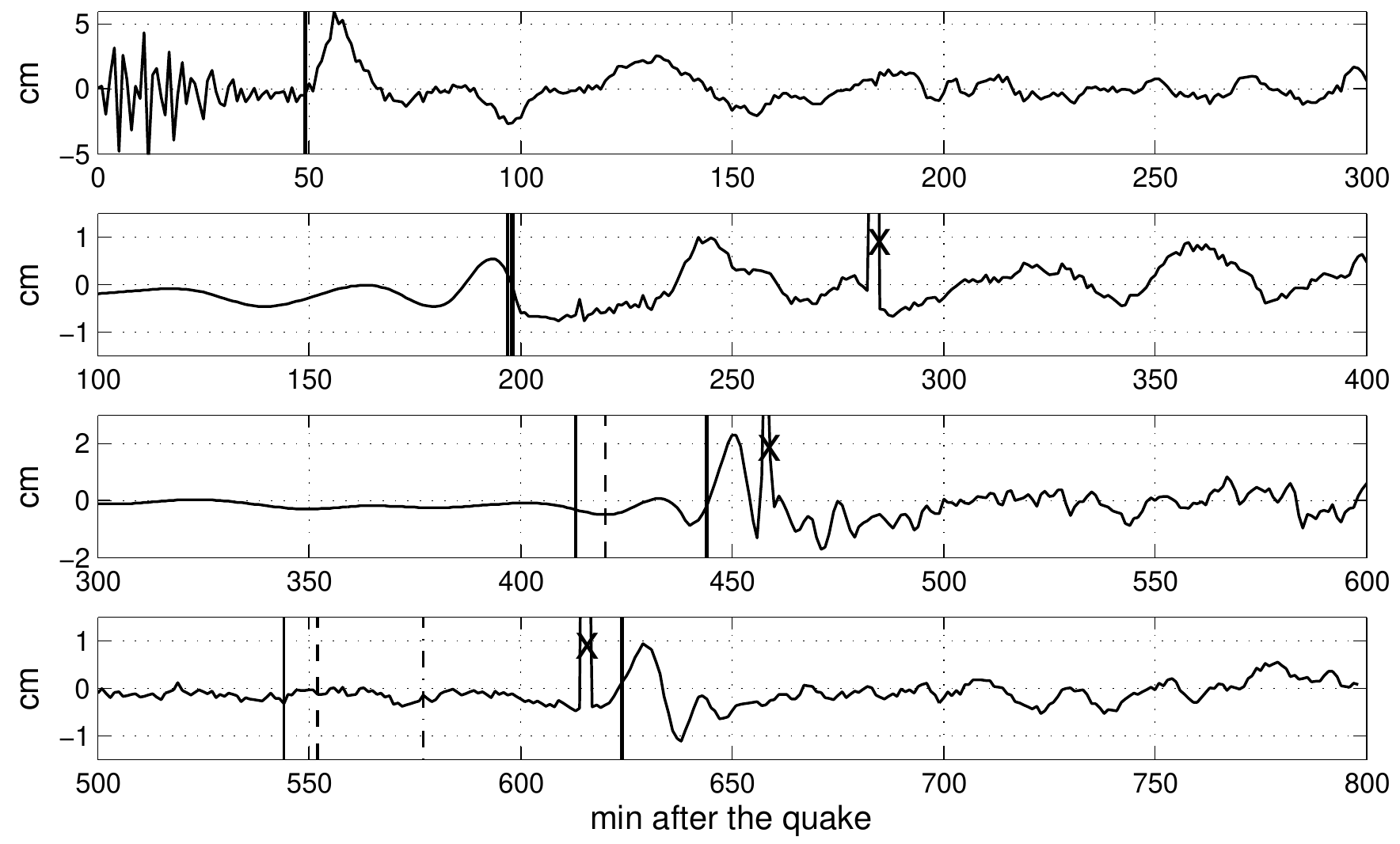} }  
	\caption{
De-tided records of the Kuril 01/2007 tsunami at DART buoys (top to bottom) 	
21413, 21414, 46413, 46408, 46419.
	}
\end{figure}

The values of $\lambda_*\tau$ and $u$ seems to suggest 
that the average speed of the front is nearly constant: 
in the interval $75\le\tau\le 640$, $\lambda_*\tau$ ranges from 11.158 to 11.680.

Calculations show that the the length of the wave  of maximum height equals to
$$
l =L_*|f|(\tan\,\theta_f-\tan\,\theta_r)\approx 5L_*|f| \,\,\,\,\,\,\hbox{for}\,\,\,\,\, 30<\tau<640.     \eqno(3.5)
	$$
\pagebreak

{\bf 4. Gauge records and data extracted from the records} 
\bigskip

Data of  2006 Kuril 
and 2007 Kuril tsunamis generated by the earthquakes of 
the 15 November 2006 and the 13 January 2007, respectively, are used to test 
the theory.  

The data is obtained from tsunami records at DART buoys 
(DART - Deep-ocean Assessment and Reporting of Tsunami) deployed in the Pacific 
Ocean at a depth of 5000 metres. Waves arriving at a deep-ocean buoy is 
a mixture of tidal, seismic, and gravity waves which come from the tsunami 
origin itself. Records of the buoys and their locations can be found on the USA 
National Data Buoy Center public website (http://www.ndbc.noaa.gov/dart.shtml).

Figures 3 shows  the Kuril 01/2007 tsunami records de-tided 
using low-pass Buttherworth filter with 150 min cut-off. 

Data extracted from the records at the DARTs  
are summarized in Tables 2 - 3, where the travel time of the MHW, $t_*\,\hbox{min}$, 
the maximum wave-height, $H_*\,\hbox{cm}$, and
the distance between the gauge sensors and the centre (epicentre of the earthquake) of 
the wave origin, $y_*\,\hbox{km}$, are shown for each of the DARTs.

\begin{center}
TABLE 2. The 2007 Kuril tsunami: WMH as recorded at the DART buoys.

Data 1.
 \begin{footnotesize}
\begin{tabular}{|p{5mm}|p{15mm} |p{10mm} |p{10mm} |p{10mm} | p{15mm} | } 
\hline $\hbox{buoy}$ &$\hbox{DART}$ & $t_*\,\,\hbox{min.}$ & $H_*\,\,\hbox{cm} $ & 
  $y_*\,\, \hbox{km}$ & $y_*/t_*^2$ \\
\hline   1&  21413    &     120    &    4.0    &    1762  &  0.122361   \\
         2&  21414    &     120    &    5.7    &    1804  &  0.125278   \\
         3&  46413    &     154    &    5.7    &    2253  &  0.094999   \\
         4&  46408    &     200    &    4.5    &    2660  &  0.066500   \\
         5&  46419    &     438    &    1.6    &    5470  &  0.028513   \\
\hline 
\end{tabular}\\
\end{footnotesize}
\end{center}

\begin{center}
TABLE 3. The 2006 Kuril tsunami: WMH as recorded at the DART buoys. 

Data 2.
\begin{footnotesize}
\begin{tabular}{|p{5mm}|p{15mm} |p{10mm} |p{10mm} |p{10mm} | p{15mm} | } 
\hline $\hbox{buoy}$& $\hbox{DART}$ & $t_*\,\,\hbox{min.}$ & $H_*\,\,\hbox{cm} $ &   $y_*\,\, \hbox{km}$ & $y_*/t_*^2$ \\
\hline   1&  46413    &     155    &    8.5    &    2331  &  0.097024   \\
         2&  46408    &     238    &    7.5    &    2735  &  0.048284   \\

         3&  46402    &     270    &    10.0   &    3127  &  0.042894   \\
         4&  46403    &     365    &    7.5    &    3581  &  0.026879   \\
\hline 
\end{tabular}\\
\end{footnotesize}
\end{center}
\bigskip 

{\bf 5. Theoretical characteristics of the WMH } 
\bigskip

The Table 1 shows that $\tan\theta_f$ and $\lambda_*$ are monotone functions of $\tau$, so 
for given value of $\lambda_*$ the values of $\tau$ and $\tan\theta_f$ can be calculated 
using equations (2.1) or estimated using Table 1.

For each DART location, setting $\lambda_*=y_*/t_*^2$ (see Tables 2 - 3) and using Table 1, 
we obtain the results, shown in Tables 4 - 5. 
Values of $\tau,\,\,$ $\tan\theta_f(\tau),\,\,$ $\Delta(\tau)$ in the first line 
of Table 4
 are obtained with the use of $\lambda_*= 0.122361$ (in the first line of Table 2) and 
Table 1 as follows. 
We see from Table 1 that 
$0,1159<\lambda_*=0,122361<0,1263,\,\,$ $90<\tau<100,\,\,$ 
$58.000<\tan\theta_f(\tau)<65.695,\,\, $ $0,3661<\Delta(\tau)<0.3840$.

 Linear interpolation gives $\tau=93.787,\,\,$ 
$\tan\theta_f(\tau)=60.914,\,\,$ $\Delta(\tau)=0,374984$. 
The rest lines in Table 4 are obtained in the same way. 

 \begin{center}
Table 4. For Data 1: theoretical characteristics of the WMH 

at locations of DART buoys.  

\begin{footnotesize}
\begin{tabular}{|p{5mm} |p{15mm} |p{15mm} |p{15mm} | p{15mm}| } 
\hline       $\hbox{buoy}$ & $\lambda_* $ & $\tau $   & $\tan\theta_f(\tau)$ 
& $\Delta$ \\ 
\hline   1     & 0.122361 & 93.787  &  60.914  & 0.374984 \\
         2     & 0.125278 & 90.983  &  58.756  & 0.378995 \\
         3     & 0.094999 & 121.125 &  78.987  & 0.335795 \\
         4     & 0.066500 & 170.000 &  108.935 & 0.288400  \\
         5     & 0.028513 & 394.300 &  251.192 & 0.193968 \\
\hline
\end{tabular}\\
\end{footnotesize}
\end{center}
Table 5 is similar to Table 4 and obtained in similar way.

\begin{center}
Table 5. 
For Data 2:  theoretical characteristics of the WMH 

at locations of the DART buoys. 

\begin{footnotesize} 
\begin{tabular}{|p{5mm}|p{15mm} |p{15mm} |p{15mm} |p{15mm} | p{15mm}| } 
\hline $\hbox{buoy}$ &  $\lambda_* $ & $\tau $   & $\tan\theta_f(\tau)$ 
& $\Delta$ \\ 
\hline 1      & 0.097024 & 118.558  &  78.411  & 0.339047 \\
       2      & 0.048284 & 235.888  &  154.791 & 0.246741 \\
       3      & 0.042894 & 265.038  &  180.234 & 0.234534 \\
       4      & 0.026879 & 420.300  &  279.545 & 0.187836 \\
\hline
\end{tabular}\\
\end{footnotesize}
\end{center}
\bigskip

{\bf 6. Estimators for the wave model parameters}
\bigskip

For the front of the wave of maximum height (as for any zero) the following formulas hold 
$$
y_*(t_*)=-f\tan\theta_f(\tau)\cdot L_*>0,\,\,\,\,\,\,
t_*=\tau \sqrt{2|f|L_*/g}.
	$$

If at each locality $i$ data for the waves (2.1) were obtained from the records exactly, the functions 
$$
S(a)=\sum_{i}\left(y_{*i}(t_{*i})-a\cdot \tan\theta_1(\tau_i)\right)^2\,\,\,\,\,\,
\hbox{and}\,\,\,\,\,\,F(a)= \sum_{i} \left(60t_{*i}-\tau_i \sqrt{\frac{2a}{g}}\right)^2
	$$
would be equal to zero at $\,a=|f_*|$, where $|f_*|=|f|L_*$.  

But using equations (2.1) and values 
of $y_{*i},\,$ $t_{*i},\,$ $H_{*i}\,$ (obtained with errors) we estimate 
the value of $a$ by minimizing $S(a)$ or $F(a)$.

Minimum value of $S(a)$ occurs at
$$
a=|f_{*1}|=\frac{ \sum_{i} {y_{*i}\tan\theta_{1i}}  }
{\sum_{i} {\tan^2\theta_{1i}}}\,\,\hbox{km},                    \eqno(6.1)      
	$$
while minimum value of $F(a)$ is reached at
$$
a=|f_{*2}|=\left(\frac{\sum_i {t_{*i}\tau_i}}{\sum_i {\tau^2_i}}\cdot 60\right)^2 4.9\,\,\hbox{m}.                                                 \eqno(6.2)      
	$$
In the Table 1, the quantity $\Delta(\tau)=H(\tau)\sqrt{2|f|}$ is independent 
of $f$ ($cH(\tau)$ is the height of the WMH). 
If the data in Tables 2 - 4 were measured exactly at each locality $i$, the function 
$$
R(s)=\sum_i(c\Delta(\tau_i)\frac{1}{\sqrt{2|f|}}\,L_* -H_{*i})^2=
\sum_i\left(\Delta(\tau_i)s-H_{*i}\right)^2,
\,\,\,\,\,\,s=\frac{c}{\sqrt{2a}}\,L_*^{3/2}
	$$
would be equal to zero at true value of $s$.

At actual values of $H_{*i}$ the value of $s$ is estimated by minimizing  
$R(s)$, which leads to the estimator 
$$
c^2L_*^3=2a\eta^2,\,\,\,\,\,\,\eta=\frac{B}{A},\,\,\,\,\,\,  
A=\sum_i \Delta^2(\tau_i),\,\,\,\,\,\,
B=\sum_i \Delta_iH_{*i},                        \eqno(6.3)      
	$$
where $\eta$ is independent of $c$ and $f$, $a$ is estimated by (6.1) or (6.2).

For Data 1, Tables 2 and 4, estimators (6.1), (6.2), and (6.3)  give 
$$
a_1=23.413\,\,\,\hbox{km},\,\,\,\,\,\,a_2=23140\,\,\,\hbox{m},\,\,\,\,\,\,
\eta=13.8\,\text{cm}.
	$$
Tables 3 and 5 (for Data 2) lead to the values 
$$
a_1=15.424\,\,\,\hbox{km},\,\,\,\,\,\,a_2=15785\,\,\,\hbox{m},\,\,\,\,\,\,
\eta=54.8\,\text{cm}.
	$$

{\bf 7. Estimation of the wave origin parameters}
\bigskip

{\bf 7.1. Dimensions of the wave origin}
\bigskip

Here, the water surface above the disturbed body of water is referred to as the wave
origin. 

By model (2.1), at the moment $t=0$ the free surface has not yet been displaced 
from its mean level (the horizontal plane $x=0$), but the velocity field has 
already become different from zero. 

Calculations show that during some time interval, say $0\le\tau\le\tau_1$, 
a water hill in the form of a rounded solitary elevation symmetric with respect to
the vertical $x$-axis is appearing on the water surface (Fig. 2). 
For the model (2.1), the height of the hill increases and reaches the maximum 
at $\tau=\tau_1=0.872$. 
On the interval $0\le\tau\le\tau_1$ there is only one zero 
$\theta(\tau)$ in the water surface (at $y>0$), and the zero is almost 
immovable. Only at $\tau>\tau_1$ the heap of water begins to spread out 
and to turn into a wave group, which runs away from the wave origin. 
The time interval $0\le\tau\le\tau_1$ is referred 
to as the interval of formation of the wave origin. 

The quantity $a=|f|L_*$ is the characteristic horizontal scale of the wave origin, 
so we determine the effective length, $l_{ef}$, of the origin by $l_{ef}=ka$ The body of water 
in the region $|y|<l_{ef}$ may be referred to as effective wave origin. 

The value of $k$ may be assigned 
to meet different conditions; for instance, at $t=0$ the energy of the water outside 
the effective origin does not exceed 5\% of the total wave energy or 
the magnitude of 
the sea surface displacement at the margin of the effective wave origin does 
not exceed a given value.

It is found from (2.1), that the maximum free surface displacement on the boundary of the wave originating area is given by 
$$
W_b=\frac{1}{\sqrt{2|f|}}\,T_1(\theta_b,\tau_1),\,\,\,\,\,\tau_1=0.842.
	$$

The quantity $-h_b=\sqrt{2|f|}\,W_b$ is independent of $f$. The values of the quantity corresponding to different values of $k=\tan\theta_b$ are shown in Table 6.

\begin{footnotesize}
\begin{center} Table 6. Values of $-h_b$ corresponding to some values of $k$.
\begin{tabular}{|p{10mm} |p{10mm} |p{10mm} |p{10mm} |p{10mm}| p{10mm}| p{10mm}| p{10mm}|} 
\hline 
$k$    & 3.0     & 3.5     & 4.0     & 4.5     & 5.0     & 5.5     & 6.0     \\ 
$-h_b$ & 0.1137  & 0.0953  & 0.0795  & 0.0666  & 0.0563  & 0.0480  & 0.0413 \\
\hline
\hline
$k$    & 6.5     & 7.0     & 8.0     & 9.0     & 10.0    & 11.0    & 12.0     \\ 
$-h_b$ & 0.0358  & 0.0314  & 0.0245  & 0.0197  & 0.0161  & 0.0134  & 0.0113  \\
\hline
\end{tabular}\\
\end{center}
\end{footnotesize}

For the model (2.1), $W_b<0$ at $k> 1.3$ (Fig. 2).
\bigskip

{\bf 7.2. Water elevation in the wave origin.}
\bigskip

Calculations  show that $\max \Delta(\tau)=\Delta(0.842)=1.036\,$ 
(the maximum is reached at $\theta=0$). This means that 
the maximum of the water elevation, $h$, in the wave origin is given as
$$
h=1,036cL_*\frac{1}{\sqrt{2|f|}}=1.036\eta \,\,\text{cm}.             \eqno(7.1)     
	$$
where $\eta$ is given by (6.3).

Formulas (6.3) and (7.1) lead to the following estimates:  
$$
 \eta_1=13.8\,\,\,\text{cm}, \,\,\,h_1=14.3\,\,\hbox{cm};
\,\,\,\,\,\,\eta_2=54.8\,\,\,\text{cm},\,\,\, h_2=56.7\,\,\hbox{cm},                          
                                                           \eqno(7.2)
	$$ 
where the subscripts 1 and 2 correspond to the Data 1 (the 2007 Kuril tsunami) and  
Data 2 (the 2006 Kuril tsunami) respectively.

The ratio $\Delta_b/1.036$ is the ratio of water surface displacement on the boudary 
of the wave origin to the maximum water displacement in the wave origin.
\bigskip

{\bf 7.3. Duration of wave origin formation.}
\bigskip

Duration $t$ of the wave origin formation is estimated 
by ($g=9.8\,\,\hbox{m/s}^2$, $a$ is measured in metres)
$$
t=0,842\sqrt{2a/g},                                 
	$$
which gives  $t=58\,\hbox{s}$ for Data 1, and $t=47\,\hbox{s}$ for Data 2. 
\bigskip

{\bf 7.4. Estimation of the waves energy}
\bigskip

It is supposed above, initially still water is set in motion at $t=0$
by an impulsive force. Kinetic energy supplied to the water by the force can be estimated by integral (involving velocity potential 
and its normal derivative) evaluated over the sea surface [3]

The energy supplied to the vertical layer of the water 
between two parallel planes at the distance $a=|f|L_*$ apart is given by   
$$
E = \frac{1}{8}\pi |f| 
c^2\cdot\gamma gL_*^4\,\,\,\hbox{J}=\frac{1}{4}\,\pi\gamma g a^2\eta^2 \,\,\,\hbox{J}.                                  \eqno(7.3)
	$$
With the values of $a$ and $\eta$ obtained in sections 5 and 7.2, we find the energy estimates as 
$$
E_1=5.03\cdot 10^{10}\,\,\text{J},
\,\,\,\,\,\,\,E_2=1.87\cdot 10^{11}\,\,\text{J},\,\,\,\,\,\,E_2=3.72 E_1     
                                              \eqno(7.4)
	$$
The subscripts 1 and 2 correspond to the Data 1 and Data 2 respectively. 

{\it Estimates (7.4) are in line with the second quotation from [1] given above}.
\bigskip

{\bf 8. Theoretical estimation of waves parameters}
\bigskip

{\bf 8.1. Estimation of the WMH length}
\bigskip

By (3.5) the WMH have the lengths  $l\approx5a$
$$
l_1\approx 117\,\,\,\,\hbox{km},\,\,\,\,\,\,
l_2\approx 76\,\,\,\,\hbox{km}
	$$
obtained from Data 1 and 2 respectively.  

The smaller value of $a$, the smaller effective size $l_{ef}=ka$ of the wave origin and wave length, the higher dominant frequency of the WMH. 

{\it These results are in line with the first quotation from [1] given above}.
\bigskip

{\bf 8.2. Speed of the wave of maximum height.} 
\bigskip

For the 2007 Kuril tsunami, using data of Table 2, we find that actual values of 
average speed of the front of WMH (during time interval $0-t_*$) 
arrived at the buoys locations are 
$$
V_{*1}=\frac{1762\cdot 60}{120}=881\,\hbox{km/h},\,\,\,\,\,\,
V_{*2}=902\,\hbox{km/h},
	$$
$$
V_{*3}=878\,\hbox{km/h},\,\,\,\,\,\,V_{*4}=798\,\hbox{km/h},\,\,\,\,\,\,
V_{*5}=749\,\hbox{km/h}
	$$
with arithmetic mean $838\,\text{km/h}$.
\bigskip

We introduce the notation $a\text{(m)}=20$ which means that the nuumber 20 is nondimensionl while $a$ is measured in metres: $a=20$ m.

The travel time $t_*$ is estimated as 
$$
t_*=\frac{\tau}{60}\cdot\sqrt{\frac{2a\text{(m)}}{9.8}}\,\,\hbox{min}.                         
                                            \eqno(8.1)    
	$$
By (3.2) and (8.1) the group velocity of the packet (2.1) is
$$
V\text{km/min}=\frac{y_*}{t_*}=\lambda_* t_*=\lambda_* \tau\, \text{km/min}^2 \cdot\frac{1}{60}\,
\sqrt{\frac{2a\text{(m)}}{9.8}}\,\,\hbox{min}.
	$$
Substituting $a=l/5 \,l\,\,\,\lambda_* \tau\approx 11.3$ (from Table 1) we obtain the relation between the group velocity and the length of the wave of maximum height as
$$
V_*\approx 11.3\,\frac{1}{\sqrt{4.9\cdot 3.6\cdot 5}}\, \sqrt{l\text{(km)}}\,\,\text{km/min}
	$$
or
$$
V_*=1.22\,\, \sqrt{l\text{(km)}}\,\,\,\text{km/min}.  \eqno(8.2)
	$$
At $l=117\,\text{km}$ formula (8.2) gives 
$V_*=13.20\,\text{km/min}=792\,\text{km/h}$.

Instantaneous speed of the wave of maximum height  is estimated as 
$$
v_*=uT_*=u\sqrt{2a\,g}=u\sqrt{17.6\,a\text{(m)}}\,\,\text{m/s}.
	$$
where $a$ is measured in metres, $g=9.8\,\text{m/s}^2$.

When $a$ is measured in kilometres and time in minuts  we get 
$$
v_*=uT_*=u\,\sqrt{17.6\cdot 3.6}\,\,\sqrt{a\text{(km)}}\,\,\text{km/min}.
	$$
Substituting $a=l/5$ and taking $u=(0.50+0.76)/2$  (from Table 1) we obtain
$$
v_*=2.24\,\sqrt{l\text{(km)}}\,\,\text{km/min}.       \eqno(8.3)
	$$
The ratio of the wave group speed to the instantaneous speed of the wave of maximum height is
$$
\frac{V_*}{v_*}=\frac{1.22}{2.24}=0.54
	$$
at any length of WMH.

Phase velocity $c$ of harmonic waves on deep water and their wave length are related as 
 $$
c=\sqrt{\frac{gl}{2\pi}} \text{m/s}=
\sqrt{\frac{9.8\cdot3.6}{2\pi}}\,\,\sqrt{l\,\,\text{(km)}}\,\text{(km/min)}=
2.37\,\sqrt{l\,\,\text{(km) }}\,\text{(km/min)}.
	$$
The wave group speed of packet (2.1) is a half of the phase velocity of harmonic wave when the packet's wave of maximum height and the harmonic wave have equal lengths: 
$\frac{V}{c}=\frac{1.17}{2.37}=0.493$.

{\it In passing.}
By the linear theory, speed $c$ of harmonic waves, their wavelength $l$, 
and $d$, the uniform depth of the water, are related as follows
$$
c^2=\frac{gl}{2\pi}\tanh\frac{2\pi d}{l}.
	$$
For Data 1 the length of the wave of maximum height is 117 km, its speed 
$V=792 \,\,\,\hbox {km/h}=220\,\,\,\hbox{m/s}$. 

At these values we get 
$$
\tanh\frac{2\pi d}{l}=0,259,\,\,\,\,\,\,
\frac{2\pi d}{l}=0,27,\,\,\,\,\,\,d=5030  \,\,\,\hbox{m},
	$$
which is in line with the depth of the Pacific ocean.
\bigskip

{\bf 8.3. Distribution of the wave heights in the packet (2.1)} 
\bigskip

The height of the wave of number $k$ in the packet (2.1) is estimated as 
$$
H^*=|x_{max}-x_{min}|=c\frac{\Delta_k \cdot L_*}{\sqrt{2|f|}}=
c\Delta_k\sqrt{\frac{L_*^3}{2|f_*|}}=\Delta_k(\tau) \eta.       \eqno(8.4)    
	$$
For the wave of MH the value of $\Delta_k(\tau)=\Delta(\tau)$ is shown 
in Table 1.

With time in the packet a central part is developing
which containes waves of nearly equal heights. One can see the central part in the interval $250<y<400$ at $\tau=50$ and in the interval $500<y<700$ at $\tau=100$ (fig. 2). With time the 
length of the central part increases. In the central part $\Delta_k(\tau)\approx \Delta(\tau)$ shown in Table 1.

It was found above from data 1 that $\eta_1=13.8$ cm and 
 $\eta_2=54.8$ cm from data 2. By (8.4) the maximum wave heights 
are estimated as $H_1^*=13.8 \Delta(\tau)$ cm for data of 2007 and 
$H_2^*=54.8 \Delta(\tau)$ cm for data of 2006. The ratio of the maximum wave 
heights $H_2^*/H_1^*=3.96$, so the ratio of wave heights in the central parts 
of the tsunami waves approximately equals 3.96. 

The ratio is equal 3.96, if the wave heights are taken at the same value of $\tau$. Under this condition, from  (8.1) and (3.1) we get 
$$
\frac{t_{*2}}{t_{*1}}=\sqrt{\frac{a_2}{a_1}}=\sqrt{\frac{y_{*2}}{y_{*1}}}.
                                          \eqno(8.5)
	$$
where the subscripts 1 and 2 correspond to Data 1 (the 2007 Kuril tsunami) and  
Data 2 (the 2006 Kuril tsunami) respectively. 

It follows from (3.5) and (8.5) that
$$
\frac{y_*}{a}=\frac{5y_*}{l},
	$$
$$
\frac{y_{*2}}{l_2}=\frac{y_{*1}}{l_1}=n,\,\,\,\,\,\,\,\frac{t_{*2}}{t_{*1}}=\sqrt{\frac{a_2}{a_1}}.    \eqno(8.6)
	$$
When the conditions (8.6) are  satisfied approximately  the ratio of the maximum wave heights $\frac{H_2^*}{H_1^*}$ of the Kuril tsunamis may be greater or less than 3.96.

It seems in [1] the calculations were perfomed  at proper values of time and at proper numbers of waves (of crests).

{\it This result agrees with the third quotation from [1].} 
\bigskip

{\bf 9. Theoretical forecast for the waves recorded.}
\bigskip

In Table 2  the darts are arraned from top to bottom according to their arrival time.

Consider the situation when the first two lines in Table 2 are known, but the records of the next three bouys are not obtained yet. 

Bellow, starting from the first two lines of Table 2, a line of forecasts of the WMH arrival time and amplitude at the next three buoys is produced corresponding to the timeline of the DART records.

For each of the next three buoys $\tan\theta_f=y_*/a$ is calculated, which is then used 
to  locate the values of $\tau$ and $\Delta$ between two appropriate consecutive values 
from Table 1. 

Then, for each buoy, the travel time of the WMH and its height at the locations 
of the buoys are estimated by (8.1) and (8.4)
\bigskip

{\bf 9.1. The forecast based on Data 1}
\bigskip

{\it The forecast based on two DART records.}

The first two lines of Table 2,
the first two lines of Table 4, 
and formulas (6.1) and (6.3) give 
$$
a=|f_{*1}|=29.783 \,\,\hbox{km},\,\,A=0,28425,\,\,B=3,066208\,\,\text{cm}
\,\,\eta=12.877\hbox{cm}
	$$

{\it For the buoy 3} we obtain 
$$
\tan\theta_f=\frac{y_*}{a}=\frac{2253}{29.783}=75.648.
	$$
From Table 1 we see that $70.734<75.648<76.421$.
This gives intervals for $\tau$ and $\Delta$:
$$
110<\tau<120,\,\,\,\,0.3371<\Delta<0.3607
	$$
Linear interpolation gives the estimates $\tau=116,3921,\,\,\,\,\Delta=0,342007$.

The travel time of the WMH and its height at the location 
of the buoy 3 are estimated by (8.1) and (8.2)
($a$ is taken in metres) as 
$$
t^*=151\,\,\text{min},\,\,\,\,\,\,H^*=4.4\,\,\text{cm}
	$$
{\t For buos 4 and 5} we get
$$
\tan\theta_f=\frac{y_*}{a}=\frac{2660}{29.783}=89.313,\,\,\,\,\,\,
\tan\theta_f=\frac{y_*}{a}=\frac{5470}{29.783}=183.663
	$$
The forcast for the buoys 4 and 5 is obtain on the same lines as for buoy 3.	

\begin{center}
Table 7. For Data 1: the forecast based on two DART records.
\begin{footnotesize}
\begin{tabular}{|p{10mm} |p{10mm} |p{10mm} |p{10mm} | p{10mm} |p{10mm} |}  
\hline $\hbox{Buoy}$ & ${\tan}\theta_f $ & $\tau $   & $\Delta$  
&$t^*\,\,\hbox{min}$ & $H^*\,\,\hbox{cm}$ \\ 
\hline  
         3     & 75.785  & 116.570  &  0.341764 & 151  & 4.4  \\           
         4     & 89.313  & 140.987  &  0.314194 & 183  & 3.4  \\
         5     & 183.663 & 285.338  &  0.223815 & 370  & 2.9  \\
\hline
\end{tabular}\\
\end{footnotesize}
\end{center}

{\it The forecast based on three DART records.} The following forecast is based on 
the measurements on DART buoys 21413, 21414, and 46413. 

The first three lines of Table 2, the first three lines of Table 4,  
and formulas (6.1) and (6.3) give 
$$
a=|f_{*1}|=29.197 \,\,\,\hbox{km},\,\,\,\,\,\,\eta=14.04\,\hbox{cm}.
	$$
\begin{center}
Table 8. For Data 1: the forecast based on three DART records.
\begin{footnotesize}
\begin{tabular}{|p{10mm} |p{10mm} |p{10mm} |p{10mm} | p{10mm} |p{10mm} |}  
\hline $\hbox{Buoy}$ & ${\tan}\theta_f $ & $\tau $   & $\Delta$  
&$t^*\,\,\hbox{min}$ & $H^*\,\,\hbox{cm}$ \\ 
\hline  
          4     & 91.116  & 143.305  &  0,311829 & 184 & 4.4  \\
          5     & 183.305 & 291.131  &  0.223815 & 374 & 3.1  \\

\hline
\hline
\end{tabular}\\
\end{footnotesize}
\end{center}
\bigskip

{\it The forecast based on four DART records.} The following forecast is based on 
the measurements on DART buoys 21413, 21414, 46413, and 46408. 

The first four lines of Table 2, the first four lines of Table 4,  
and formulas (6.1) and (6.3) give 
$$
a=|f_{*1}|=26,952 \,\,\,\hbox{km},\,\,\,\,\,\,\eta=10.926\,\hbox{cm}.
	$$
\begin{center}
Table 9. For Data 1: the forecast based on four DART records.
\begin{footnotesize}
\begin{tabular}{|p{10mm} |p{10mm} |p{10mm} |p{10mm} | p{10mm} |p{10mm} |}  
\hline $\hbox{Buoy}$ & ${\tan}\theta_f $ & $\tau $   & $\Delta$  
&$t^*\,\,\hbox{min}$ & $H^*\,\,\hbox{cm}$ \\ 
\hline  
          5     & 202.949 & 316.371  &  0,215841 & 390 & 2.1  \\
\hline
\end{tabular}\\
\end{footnotesize}
\end{center}

For Data 1, the results are summarized in Tables 10  where 
for each of the mentioned DART buoy's locations the forecast of travel
time $t_k^*$ (in minutes) of the wave of maximum height and its height $H_k^*$ 
(in centimetres) are shown; the subscript $k$ shows that the forecast is based 
on measurements obtained from $k$ buoys. 

The actual travel time $t_*$ and height $H_*$ are repeated from Table 1. 

\begin{center}
Table 10.  For Data 1: the forecasts based on the model (2.1).
\begin{footnotesize} 
\begin{tabular}{|p{5mm} |p{6mm} |p{6mm}|p{6mm} |p{6mm} | p{6mm}| p{6mm} | p{6mm} | p{6mm} | | } 
\hline $\hbox{Buoy}$  & $t_2^* $ & $t_3^* $ & $t_4^*$ & $t_*$ &$H_2^*$&$H_3^*$ & $H_4^*$ & $H_*$   \\
\hline 
    
         3   & 151 &  -  &   -  & 154 & 4.4  & -  &  -  &  5.7 \\
         4   & 183 & 184 &  -   & 200 & 3.4  &4.4 &  -  &  4,5 \\      
         5   & 343 & 374 & 390  & 438 & 2.9  &3.1 & 2.4 &  1,6 \\
\hline
\end{tabular}\\
\end{footnotesize}
\end{center}
In figure 3 vertical lines mark the arrival of the front 
of the WMH (thick solid line), its estimate with the first and second buoy records 
for the next three buoys (thin solid), its estimate with the first three buoy records 
for the next two buoys (dashed), its estimate with the four buoys for the last one 
(dashdot). Cross marks a trigger pulse (signals send by an operator).
\bigskip

{\bf 9.2.  The forecast based on Data 2}
\bigskip

The Data 2 obtained from the records are shown in Table 4. The results of forecasting are presented in Table 11.

\begin{center}
Table 11. For Data 2: the forecasts based on the model (2.1).
\begin{footnotesize}
\begin{tabular}{|p{10mm} |p{8mm} |p{8mm} |p{8mm} | p{8mm}| p{8mm} | p{8mm} | p{8mm} |
p{8mm} | p{8mm} | p{8mm} | } 
\hline $\hbox{Buoy}$ & $t_1^* $ & $t_2^* $   & $t_3^*$ & $t_*$ &$H_1^*$ & $H_2^*$ & $H_3^*$ & $H_*$  \\
\hline 
         2     & 187 & -   &  -   & 238 & 7.8 & -   & -   & 7.5 \\ 
         3     & 214 & 257 &  -   & 270 & 7.3 & 6.6 & -   & 10.0 \\
         4     & 243 & 295 &  307 & 365 & 6.9 & 6.2 & 6.8 & 7.5 \\
\hline
\end{tabular}\\
\end{footnotesize}
\end{center}
\bigskip

{\bf Discussion} 
\bigskip

1. When the above  estimates for a hypohetical tsunami are applyed to actual tsunamis one should keep in mind that the estimates may be reliable only for far-field sites since the DARTs are situated at long distances from the tsunami source.

2. Figures (7.4) underestimate the actual energy of the Kuril tsunamis as 
the energy flux depends on azimuthal direction.

Modelled energy fluxes for the Kuril tsunamis were directed southeastward from the source areas, in the direction of the Havaian Islands and Peru-Chile [1].
 while all DARTs named in the Tables 2 and 3 are situated  in the direction of Alaska. 

Also underestimation of water elevation in the wave origin has resulted from the   
  discrepancy between the directions. 

3. The figures (7.4)  may be interpreted as estimates of the wave energy radiated in a sector  containing the DARTs. That is why 
our present results shows good agreement with those obtained in [1] and
the predicted arrival times and predicted wave heights are comparable with that obtained from the records. 

4. It is unusual that long waves may be considered in the water of infinite depth. But the answer to a problem depends on its formulation.

The problem on gravitational wave on infinite water surface must involve  
conditions at infinity along the surface. In the linear theory of water waves  (no matter the water depth is finite or infinite) speed of the waves along the water surface should be   
bounded by a constant. This condition leads to solution in the form of sinusoidal waves. Sinusoidal waves require the energy supplied to the water by a source of disturbances to be infinite.

In the theory of specific wave packets the energy supplied to the water is finite at any moment of time. Consequently, periodic waves on the free surface infinite in extent (including sinusoidal and progressive Stokes waves) are 'prohibited' by this condition. 

The energy supplyed to the water by a quake is finite. This may explain 
the fact that the speed of tsunami in an open sea and group speed of the packet (2.1) are almost the same.
\bigskip

{\bf REFERENCES}
\bigskip

1. A.B.Rabinovich,  L.I.Lobcovsky at all, Near source observation and modelling of the Kuril Iseland tsunamis of 17 november 2006 and 13 January 2007. Adv.Geosci, 4,105-116, 2008/ www.adv-geosci.net/14/105/2008.

2. Mindlin, I.M., Deep-water gravity waves: nonlinear theory of wave groups.
arXiv:1406.1681v1 [physics.ao-ph], 30 p, (6 Jun 2014)

3. Miln-Thomson, L.M.,  Theoretical Hydrodynamics,  
Macmillan and Co. LTD, London, 1960.

\end{document}